\documentstyle[11pt,moriond,epsfig,fancybox,amstex]{article}

\def\Journal#1#2#3#4{{#1} {\bf #2}, #3 (#4)}

\def\NIM{\em Nucl. Instrum. Methods}

\def\NPB{{\em Nucl. Phys.} B}
\def\PLB{{\em Phys. Lett.}  B}
\def\PRL{\em Phys. Rev. Lett.}
\def\PRD{{\em Phys. Rev.} D}

\def\be{\begin{equation}}
\def\ee{\end{equation}}
\def\bea{\begin{eqnarray}}
\def\eea{\end{eqnarray}}

\def\dz{D\O\,\,}
\def\pt{$P_{\rm T}\,\,$}
\def\gvcc{GeV/$c^{2}$\,\,}
\def\gevc{GeV/$c$\,\, }

\newcommand{\met}{\mbox{${\hbox{$E$\kern-0.6em\lower-.1ex\hbox{/}}}_{\rm T}$}}

\begin{document}
\vspace*{4cm}
\title{NON SUSY SEARCHES AT THE TEVATRON}

\author{ R.Vilar Cortabitarte}

\address{Instituto de F{\'i}sica de Cantabria - CSIC/UC,\\
 Av. Los Castros s/n,  
Santander, 39005, Spain}

\maketitle\abstracts{The Fermilab Tevatron collider experiments, CDF
and \dz, have collected over 200 ${\rm pb}^{-1}$ of data at
$\sqrt{s}=1.96$ TeV since March 2002 (RunII). Both experiments have
investigated physics beyond the Standard Model; this paper reviews some
of the recent results on the searches for new phenomena, concentrating
on Z', extra dimensions, excited electrons and leptoquarks. No
signal was observed, therefore stringent limits on the signatures and
models were derived.}

\section{Introduction}
The Standard Model of particle physics (SM) has confirmed many of its
predictions which have been measured with great accuracy over the past
years. In spite of its success, there are some hints that it can not be a
complete model: the electroweak breaking symmetry mechanism is not
explained, gravity is not implemented, there are hierarchy problems,
etc. Several extensions to the SM have 
been proposed to address these issues: Extra
Dimensions (ED), Grand Unified theories (GUT), Technicolor (TC),
SuperSymmetry (SUSY), etc.  These models predict new signatures  that can be seen at the
experiments at small rates such as dilepton events, lepton plus jets,
jets plus missing transverse energy (\met), etc. CDF~\cite{CDF} and
\dz~\cite{D0} have searched for these processes using
$\approx \,200 \,{\rm pb}^{-1}$ of proton-antiproton collisions at $\sqrt{s}
\,=\,1.96$ TeV collected from March 2002 to October 2004.

\section{Z' searches}
\begin{figure}[!t]
\begin{minipage}{0.4\linewidth}
\epsfig{figure=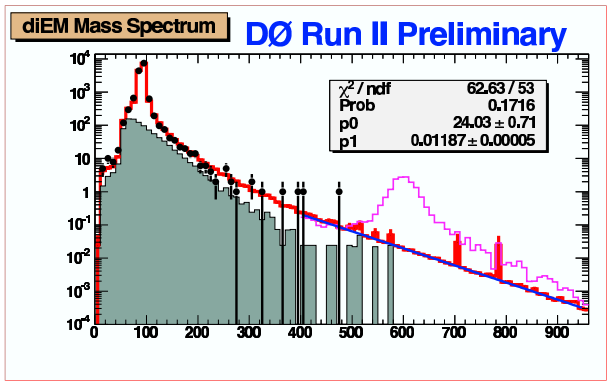,height=2.in}
\end{minipage}
\hspace*{3cm}
\begin{minipage}{0.4\linewidth}
\epsfig{figure=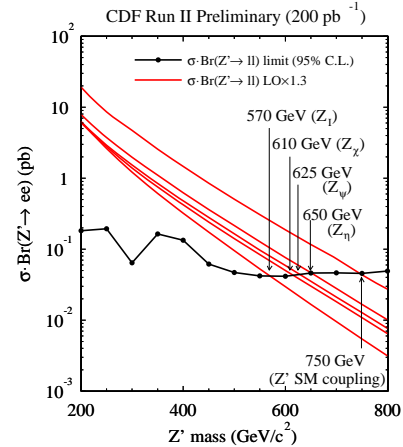, height = 2.4in}
\end{minipage}
\caption{Left:Invariant mass distribution of the two electrons using the
\dz detector. Points: data, shaded region: QCD background,
open histogram: Drell-Yan plus QCD background. Blue line: fit
to the background, magenta: Z' of 600 GeV . The fit
parameter p1 corresponds to the negative slope of the exponent
while p0 reflects the normalization.
Right: CDF 95\%C.L. upper limit for the different model of the Z'
using the electron channel.}
\label{fig:zlim}
\end{figure}

A heavy partner of the Z boson, the so-called Z' boson~\cite{zp}, is predicted in
many extensions of the SM, such as GUT, ED models and little
Higgs models among many others. It is a spin-1 object. The couplings
to the SM fermions could be SM-like or modified.  
Both experiments have searched for signal of Z'. As a reference model for experimental comparisons, a Z' with SM-like couplings, and also a model 
inspired by GUT SO(10) and $E_6$ model using the conventions on
~\cite{zp1} and ~\cite{zp2} where additional Z-bosons
originating from low energy $E_6$( $Z_{I}$, $Z_{\psi},\,Z_{\chi},\,X_{\eta}$) are used to set limits. 
The primary observable is an excess production of dilepton pairs at
large invariant masses. 
CDF and \dz have looked in the dielectron
channels using $\approx$ 200 ${\rm pb}^{-1}$ of data, requiring high
\pt electromagnetic objects in the calorimeters (\pt$>25$ \gevc
). The main background comes from Drell-Yan production and QCD processes (missidentified jets),   
and a small contribution from electroweak processes. Neither experiment
observed any deviation from expectations, as shown in
Figure~\ref{fig:zlim} (left). They set a 95\% C.L. upper limit,
Figure~\ref{fig:zlim} (right). Based on
spin-1 particle for acceptance as a function of the boson mass CDF obtained an upper limit of 750 \gvcc for the SM-like case.  
\dz has set an upper limit of 780 \gvcc based on
PYTHIA Z' simulation for acceptance and a search window optimized for
each Z' mass. The limits were set in terms of Z' to Z cross section to
reduce the  systematic uncertainties. For the $E_{6}$ GUT model, lower limits
found by CDF are 570, 610, 625 and 650 \gvcc respectively; while
\dz found 575, 640, 650 and 680 \gvcc. 

CDF has also used the muon channel, setting an upper limit of 735 \gvcc for
the SM-like model and 518, 590, 620, 650 \gvcc for the $E_6$.
 
\subsection{Little Higgs Model}
This model~\cite{lh}\,~\cite{lh2} attempts to solve the hierarchy and fine
tunning problems between the electroweak scale and the Planck
scale. An explicit model~\cite{lh3} predicts new gauge bosons coupling
to the SM fermions. The coupling, purely left-handed, are universal
and scale linearly with the mixing angle. 
CDF has re-interpreted the result of the Z' analysis under this model
assumption for both channels (e, $\mu$), excluding a 95\% C.L. region on the
parameter space of the Z' mass and the mixing angle, $cot\,\theta$.
 For $cot\,\theta$ = 1, masses are excluded up to
825(790) \gvcc for the electron (muon) channel.

\section{ Large Extra Dimensions (LED)}
The hierarchy problem has motivated a number of models beyond the SM. In
recent years, a number of models in ED have been
proposed~\cite{LED}: it might exist hidden dimensions in space of
finite size R beyond the three we sense daily. 
 In hadron-hadron collisions, real gravitons can be produced in
association with jets or photons~\cite{gled}. The graviton lives in
the ED, invisible to our world and the detectors. The exact
experimental observation is different for the various models.

\subsection{ ADD Extra Dimensions}
In the so-called ADD~\cite{add} model, SM particles are confined to a
3D-brane, while gravity propagates freely in $n$ ED, compact
spatial dimensions, which explains its apparent weakness.  The radius,
$\cal{R}$,of these ED is about 1mm for n=2 and less
than 1nm for n$>$3. From direct gravitational measurement n$<$2 is
excluded~\cite{dxd}. The graviton
is equivalent to a tower of Kaluza-Klein (KK) states, with a separation
between different states of $\cal{O}$$(10^{-4})$ due to the high energy
of the experiments, therefore the mass spectrum is a continuum. 

\begin{figure}[!t]
\begin{minipage}{0.4\linewidth}
\epsfig{figure=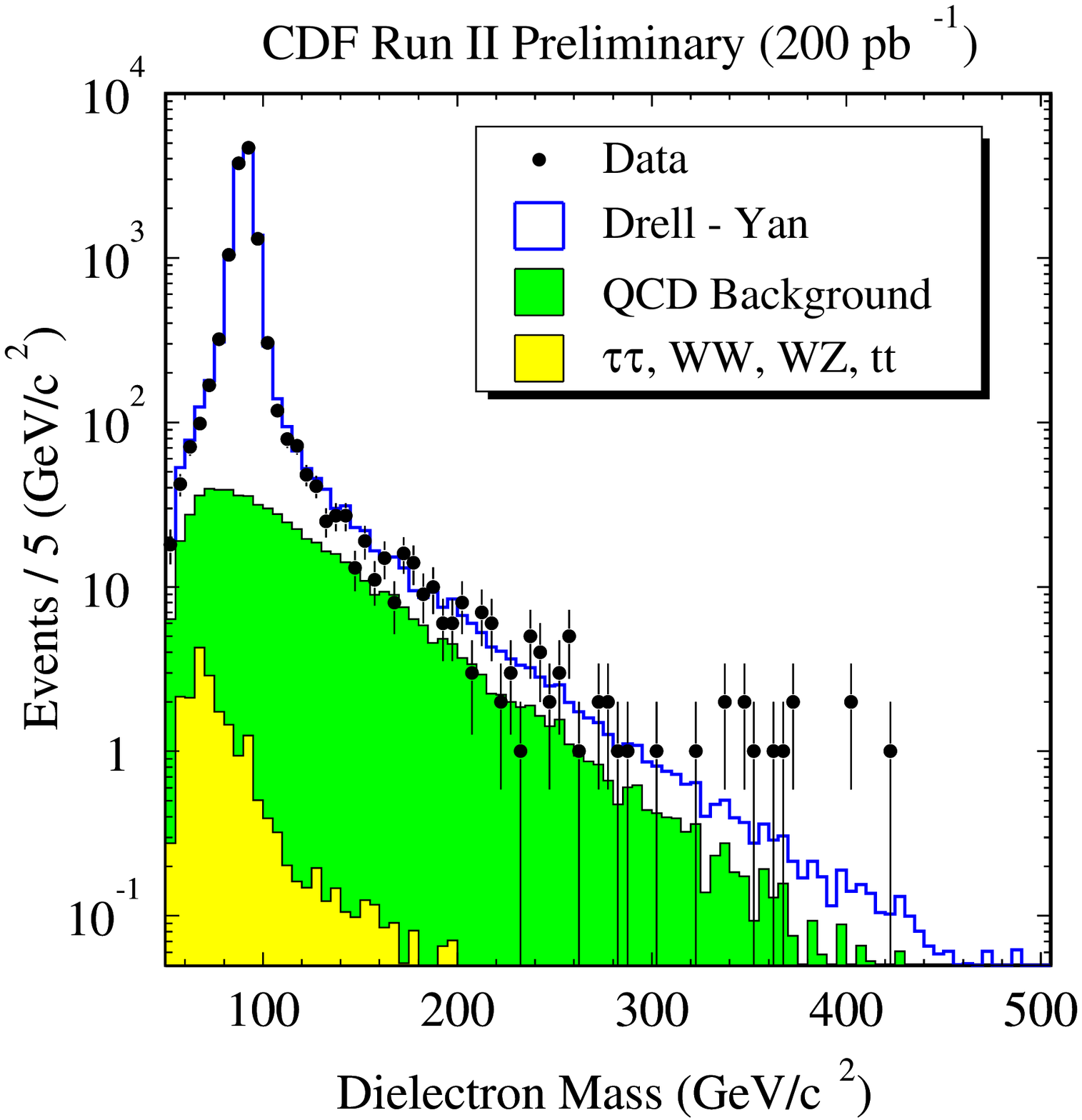,height=2.3in}
\end{minipage}
\begin{minipage}{0.4\linewidth}
\epsfig{figure=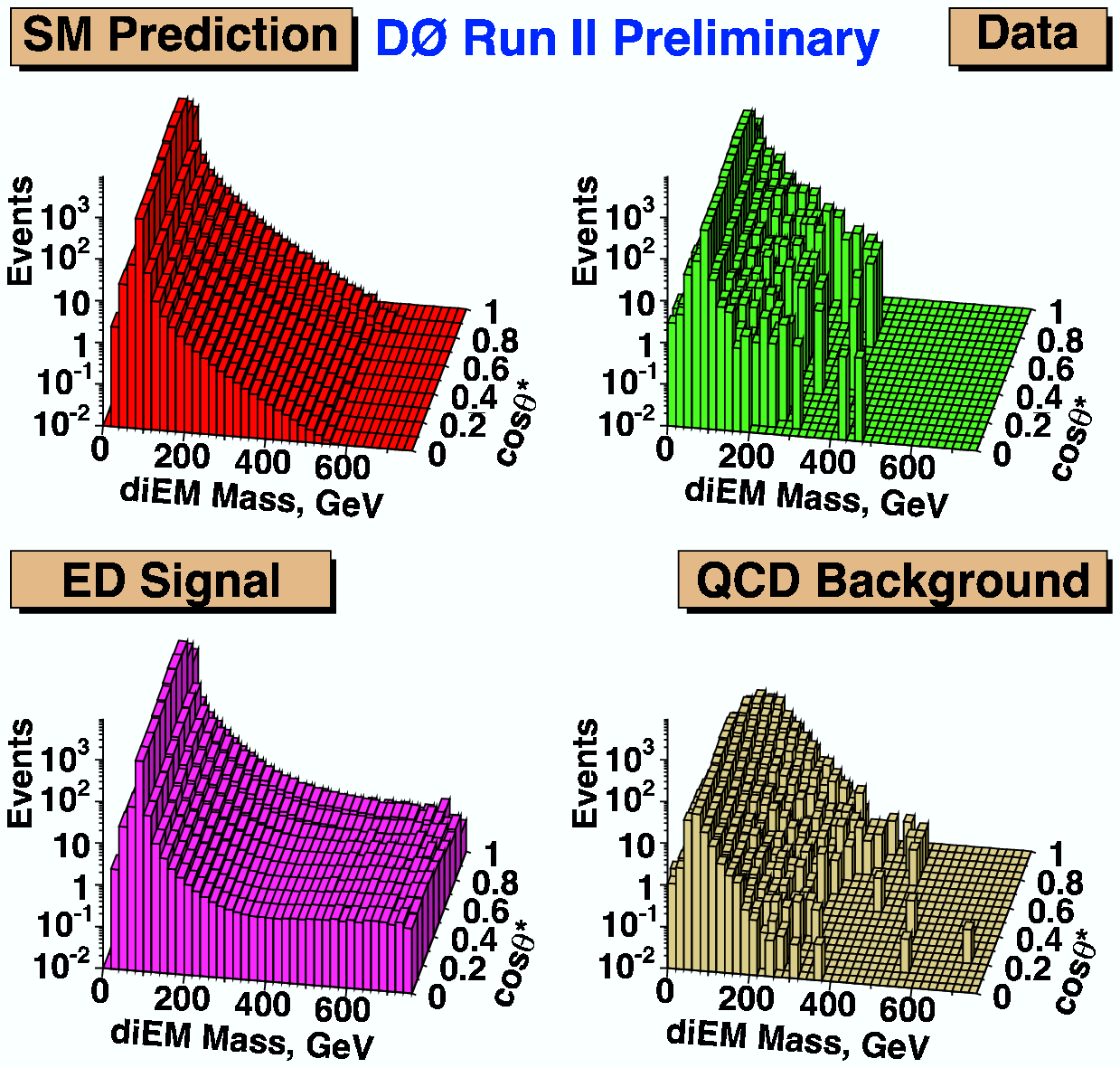,height=2.3in}
\end{minipage}
\caption{ Left side: The invariant mass distribution of the two high \pt electrons in the event.
Right side: The \dz diEM mass vs. $cos\,\theta$ distribution for
backgrounds and signal.}
\label{fig:zmass}
\end{figure}

Both experiments have searched for these
models. CDF has re-interpreted the result of the Z'
analysis where no deviation of the invariant mass of the two high
\pt electrons ($E_{\rm T}>$ 25 \gevc, and one of them with $|\eta|<$1.) with respect to
the SM predictions is observed, Figure~\ref{fig:zmass}
(left). \dz looks for high \pt ($>$ 25 \gevc and $|\eta|<$1.1) electromagnetic objects to maximize the
efficiency. \dz sets limit using a 2-D fit
to the invariant mass and the angular distributions,
Figure~\ref{fig:zmass} (right).	

CDF set an upper limit of $M_S>$1.11 TeV while the
more optimized \dz (10-15\% more sensitive) produces limit of
$M_S\, >$ 1.36 TeV. \dz combines the result with Run I setting
the most stringent limit to date, $M_S>$1.43 TeV, in the GRW convention. 

\dz looks also into the muon channel, using 100 ${\rm pb}^{-1}$
of data, setting an upper limit $M_S>$880 GeV.

With a small sample of the data, 85 ${\rm pb}^{-1}$, \dz starts a
search for the real production of the graviton, where the graviton
recoils against the jet and escapes undetected. The resulting topology
is monojet-like. The main backgrounds come from Z+jets
production and a smaller contribution from W+jets production. The
analysis~\cite{jmet} requires a high \pt leading jet ($P_{\rm T}>$150 GeV
in the central region), second jet ($P_{\rm T} <$50 ) \gevc, \met
$>$ 150 GeV,
no leptons in the event and angular separation between the jet and
\met. The expected number of events is 100$\pm$6$\pm$7, while 63 are observed. This
gives an upper limit of 84 event for an expected limit of 123.8$\pm$28 events. The current result is limited 
by the large MC and data jet energy scale uncertainties, which yields
uncertainties of 20\% for the signal efficiency and +50\%, -30\% for
the background prediction. The result is an upper limit as a function
of the number of ED, Figure~\ref{fig:jmet}.

\subsection{ Randall-Sundrum Model }
This model~\cite{RS} proposes a large curvature of the
ED to address the hierarchy problems by means of a non-factorisable
geometry in a 5 dimensional space, with a constant negative
curvature. Therefore the KK gravitons are very different to the ADD
model, the mass and couplings of each KK state is determined by the
warp factor, the spectrum of the KK states are discrete and unevenly
spaced with a coupling strength of 1/TeV for each resonance. The properties
of the Randall-Sundrum model are determined from the ratio of $k/M_{\rm pl}$, where $k$
is a scale of the order of the Planck scale and $M_{\rm pl}$ is the
effective Planck scale.

\begin{figure}[t!r]
\begin{minipage}{0.45\linewidth} 
\begin{center}
\epsfig{figure=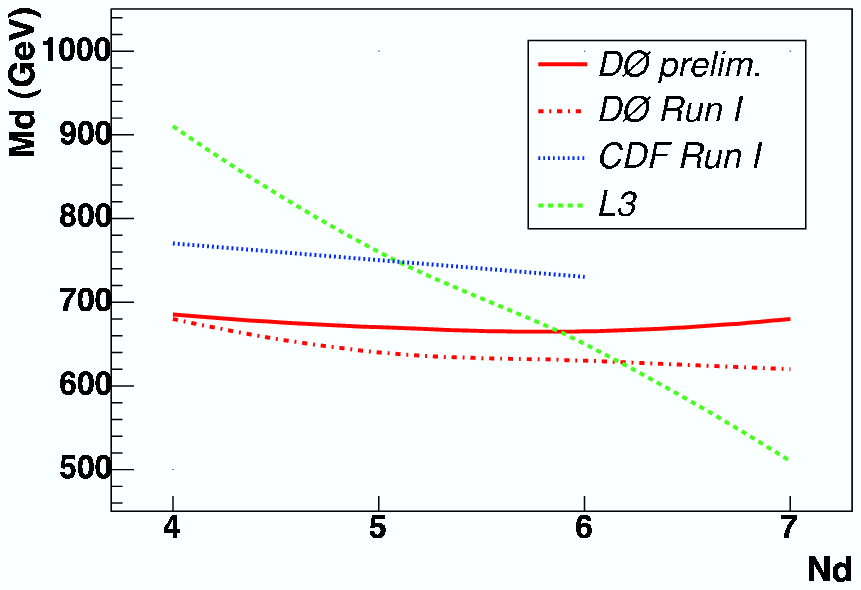,height=2.in}
\caption{ The \dz upper limit on the Planck scale as a
function of the number of dimensions}
\label{fig:jmet}
\end{center}
\end{minipage} 
\hspace*{0.5cm}
\begin{minipage}{0.4\linewidth} 
\begin{center}
\epsfig{figure=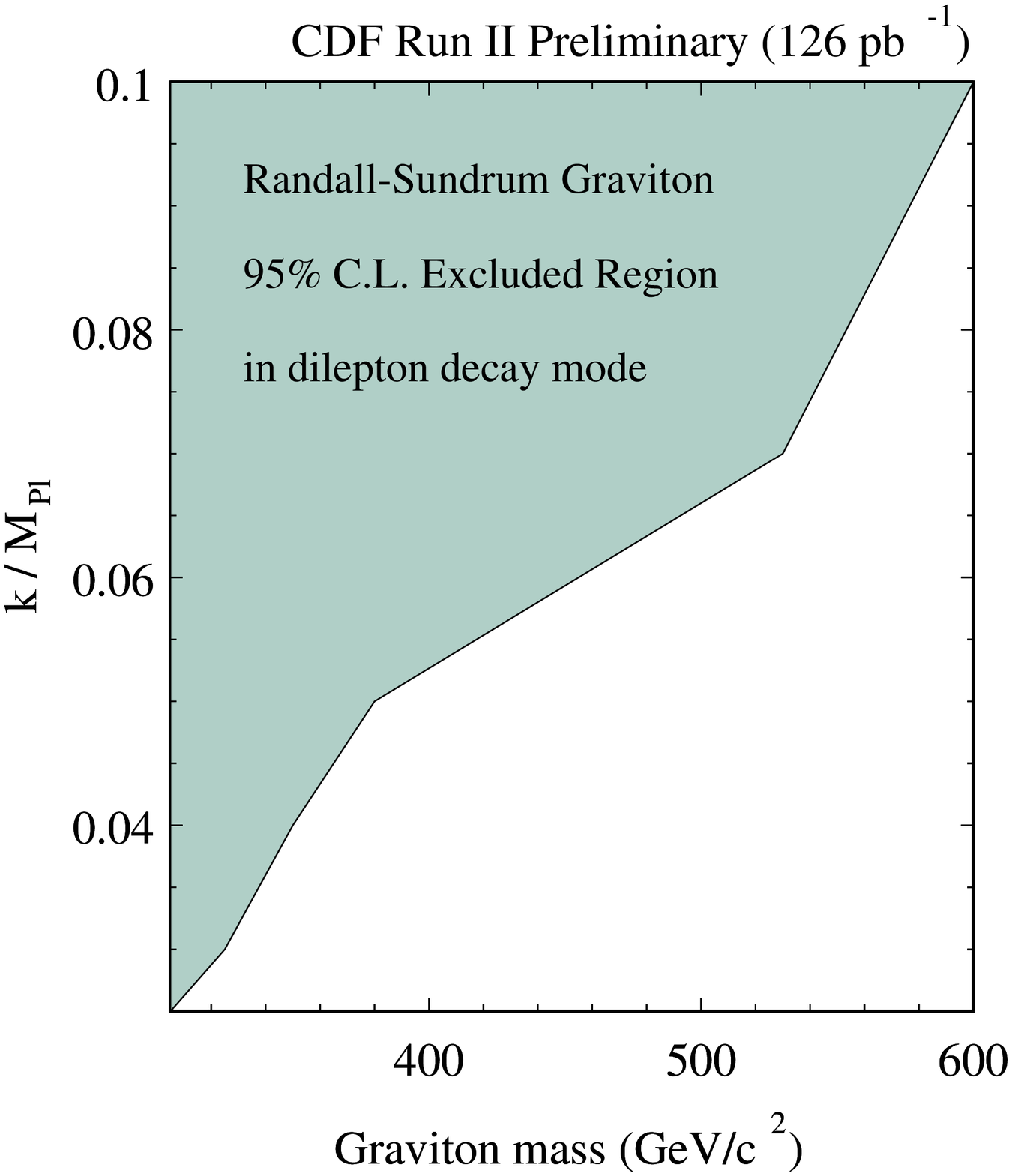,height=2.08in}
\caption{ CDF 95\% CL excluded region in the mass-coupling parameter
for the Randall-Sundrum gravitons in the dielectron }
\label{fig:RS}
\end{center}
\end{minipage}
\end{figure}

CDF has performed a search for such graviton particles in the dilepton
channels (ee and $\mu\mu$), reinterpreting the data used in the high
mass dilepton search. There is no deviation from the SM prediction so
an upper limit is set using the acceptance from a spin-2 particle and
using a likelihood fit to the invariant mass distribution. CDF can
exclude a region on the parameter space of the ratio and the graviton
mass at 95\% C.L, Figure~\ref{fig:RS}. 

\subsection{$TeV^{-1}$ ED}
In this model~\cite{tev}, matter resides on a p-brane, with chiral
fermions on the 3-brane internal to the p-brane and SM gauge bosons
propagating in all p dimensions. The compactification scale in this
model is of the order of 1/$M_{c}$. The SM gauge bosons are equivalent
to towers of Kaluza-Klein states with masses $M_n = \sqrt{M_0^2
+n^2/R^2}$. This rises mixing among the 0th and the nth-modes of the
W/Z bosons and there is a direct production and virtual exchanges of
the zeroth-states gauge bosons possible at high energies.
\dz has performed a dedicated search in the
dielectron channel, high \pt electron in the central region. This
is the first direct search and it produces a limit of $M_{\rm c}>$ 1.12 TeV.
Indirect search at LEP imply $M_{\rm c}>$ 6.6 TeV.

\section{Excited Electrons}
\begin{figure}[t!]
\begin{minipage}{0.4\linewidth}
\epsfig{figure=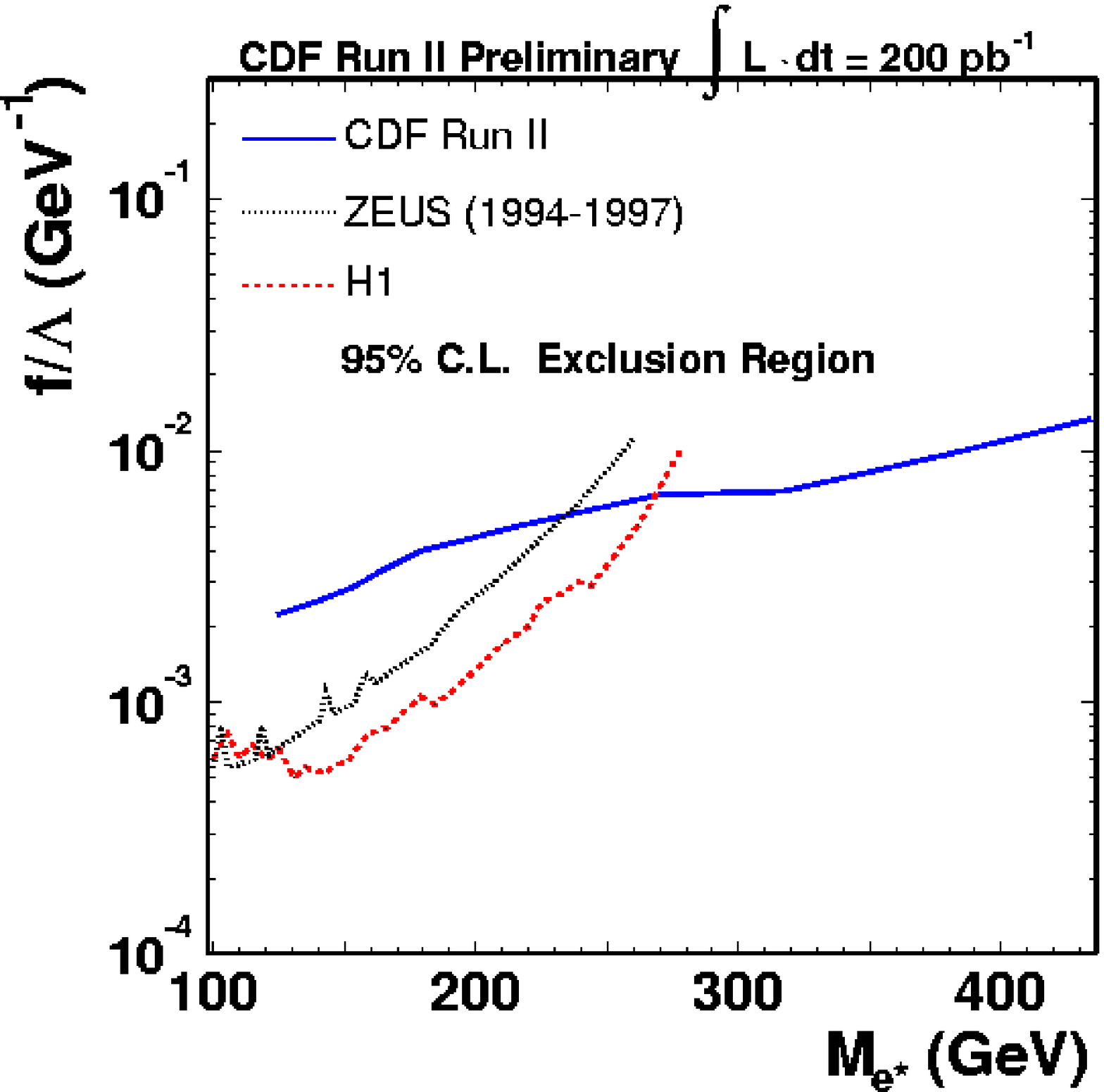,height=2.in}
\end{minipage}
\hspace*{0.5cm}
\begin{minipage}{0.4\linewidth}
\epsfig{figure=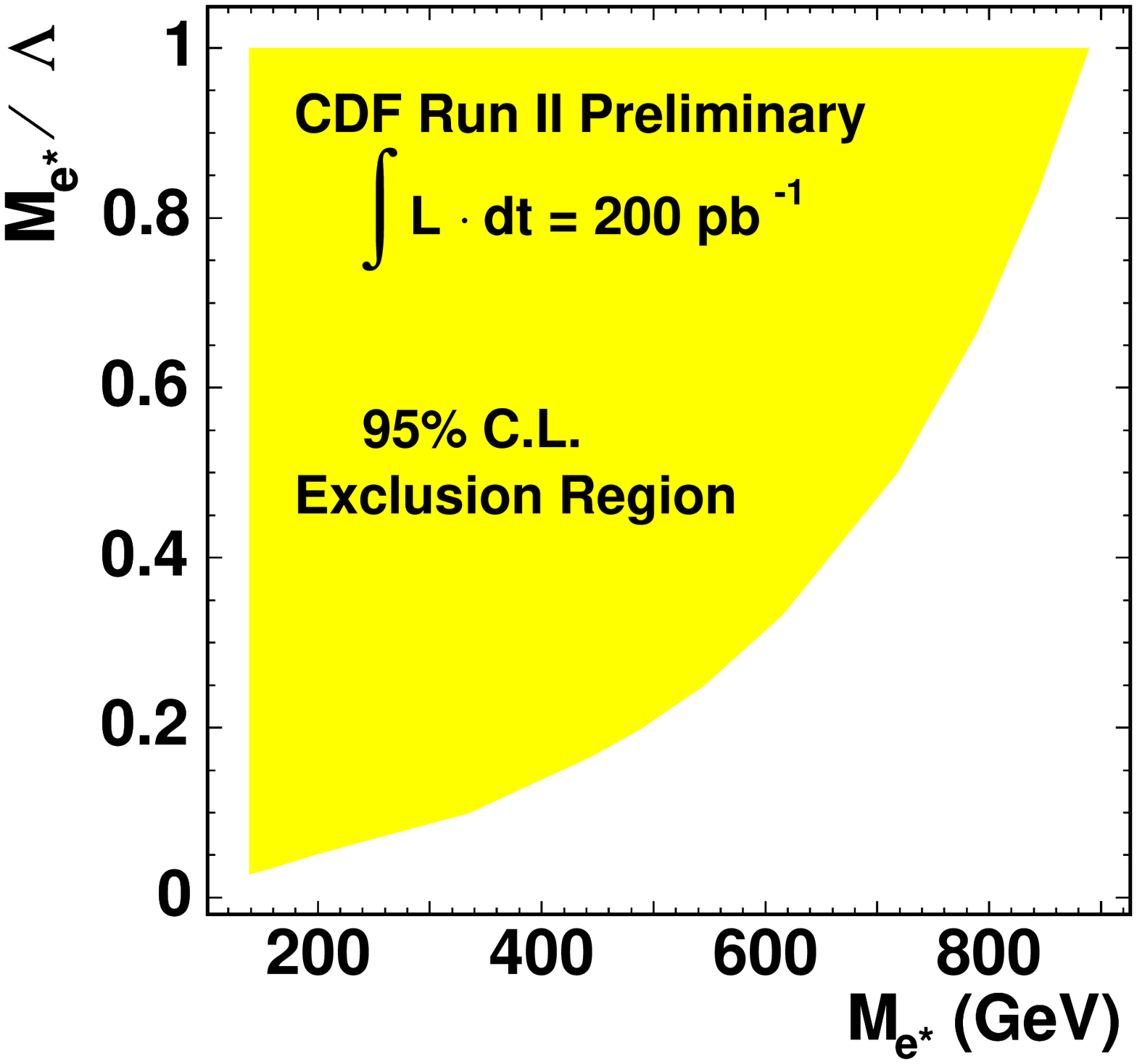,height=2.in}
\end{minipage}
\caption{ CDF 95\% CL excluded region in the mass parameter versus
$f/\lambda$ for the gauge mediated model or versus $M_{e^*}/\Lambda$
for the contact interaction model.}
\label{fig:exe}
\end{figure}

The observation of excited
particles would be a clear signal of the substructure of the
matter. At hadron colliders excited electrons, $e^*$, could be produced through
either contact interaction or gauge mediated
interactions~\cite{exe}. 
CDF has searched for $e^*$ decaying into an electron
and a photon using the 200 pb$^{-1}$. An isolated central lepton with high \pt
($>$ 20 GeV) and another EM object are requiered.
There is no observation of any deviation from the SM
prediction, Z$\gamma$ + Drell-Yan, Z+jets, WZ, QCD multijets and
$\gamma\gamma$  + jets, after applying the selection 
criteria. Therefore upper limits are set and at 95\% C.L a region in the
parameter space of the electron mass and the $M_{e*}/\Lambda$ for the
contact interaction model or $f/\lambda$ for the gauge mediated model
is excluded. The search for the contact interaction model is the first
search done at hadron colliders. Figure~\ref{fig:exe} shows the
excluded region for the two models.

\section{Leptoquarks}
Leptoquarks($LQ$) are color triplet bosons carrying both lepton and quark
quantum numbers, they can be scalar (spin = 0) or vector (spin =
1). They are predicted in many extensions of the SM such 
as GUT, Technicolor, SUSY, etc~\cite{$LQ$}. At the Tevatron, they are pair
produced mainly through gluon fusion or $\rm {q\bar{q}}$
annihilation. $LQ$ can decay either into a charge
lepton and a quark (branching ratio of $LQ$ into
charged lepton and a quark ($\beta$ = 1) or neutrino and a quark
($\beta$ = 0). Both experiments have looked for scalar $LQ$,
assuming that they only couple to quarks and leptons of the same
generation. 
\subsection{ First Generation $LQ$}
\dz and CDF have looked into the $LQ\bar{LQ }\, \rightarrow \,
e^{\pm}e^{\mp}jj$ and $LQ\bar{LQ}\, \rightarrow \,e^{\pm}\nu jj$. \dz has
use 175 pb$^{-1}$ for both channels, CDF has used 200 pb$^{-1}$ of data
for the first and 72 pb$^{-1}$ for the latter. They
required two electron with high \pt$>$25 GeV/$c$ and at
least two jets reconstructed in the calorimeter. To reduce the main
backgrounds coming from Z, Drell-Yan, $t\bar{t}$ and QCD, topological cuts
are applied. There is no deviation from the expectations. \dz finds an
upper limit on the $LQ$ mass $>$ 
238(194) GeV/$c^{2}$ for the eejj(e$\nu$jj) channel, while CDF obtains 230(166)
GeV/$c^{2}$. The results from these channels have been combined and the
derived lower mass limit as a function of the $\beta$ is shown in
Figure~\ref{fig:LQ1}.

\subsection{ Second Generation $LQ$}
CDF and \dz have done a search for the second generation $LQ$
decaying through the channel $LQ\bar{LQ} \rightarrow \mu\mu$jj. CDF has
done the analysis with 198 pb$^{-1}$ of data, while \dz uses
104 pb$^{-1}$. The events are required to have at least two 
opposite-sign muons with high \pt $>$ 25 GeV/$c$ and two
jets. To reduce the background coming from Z/Drell-Yan and $t\bar{t}$,
kinematical and topological cuts are applied. The events passing the
selection criteria are consistent with the SM expectation.
$LQ$ mass below 240 GeV/$c^{2}$ is excluded by CDF, while \dz obtains a limit of M $>$ 186 GeV/$c^{2}$

\subsection{ All Generation $LQ$}
\begin{figure}[t!]
\begin{minipage}{0.45\linewidth} 
\begin{center}
\epsfig{figure=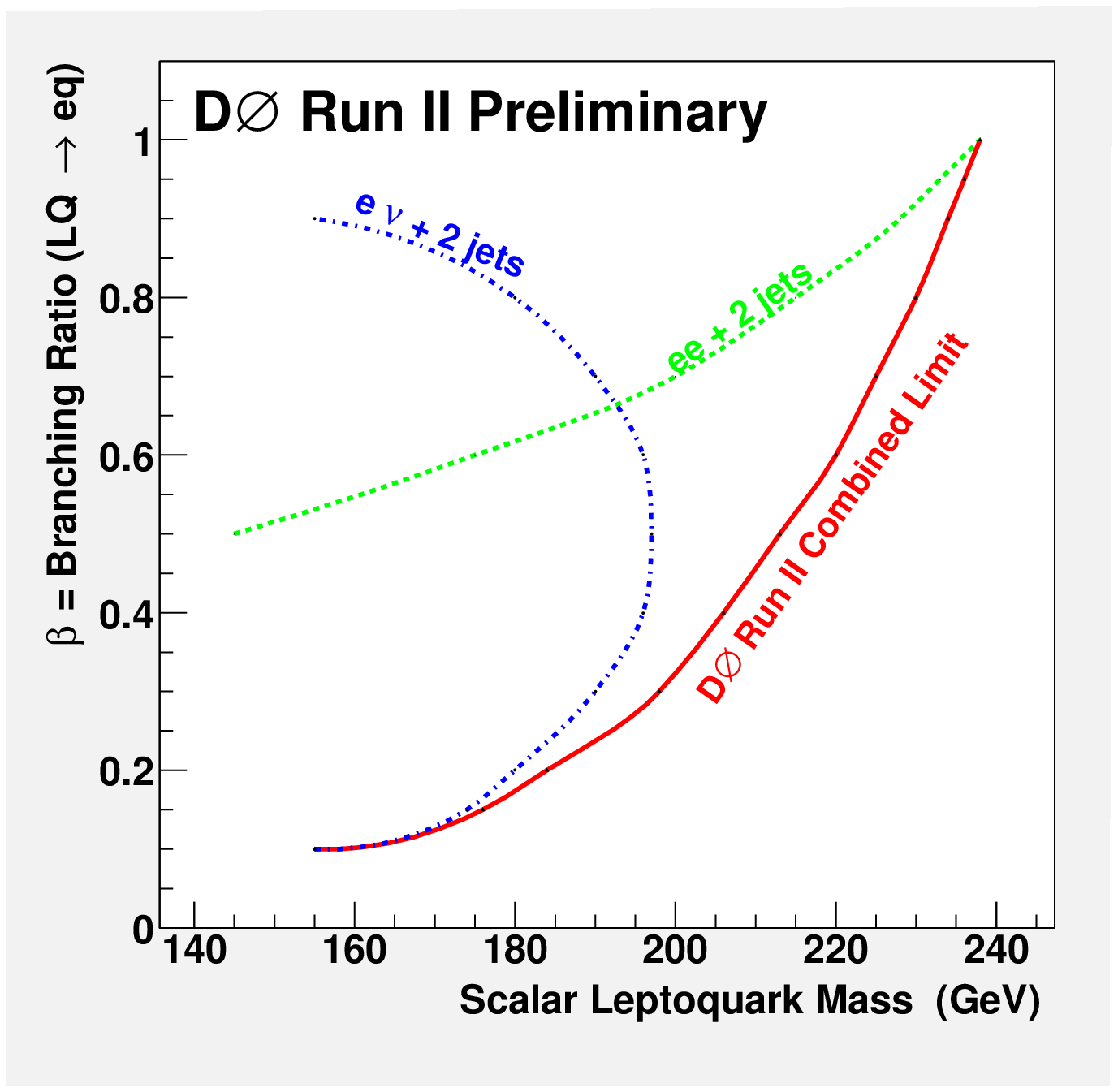,height=2.in}
\caption{ \dz 95\%C.L. lower limit on the mass of the first generation
$LQ$ as a function of $\beta$}
\label{fig:LQ1}
\end{center}
\end{minipage} 
\hspace*{0.8cm}
\begin{minipage}{0.4\linewidth} 
\begin{center}
\epsfig{figure=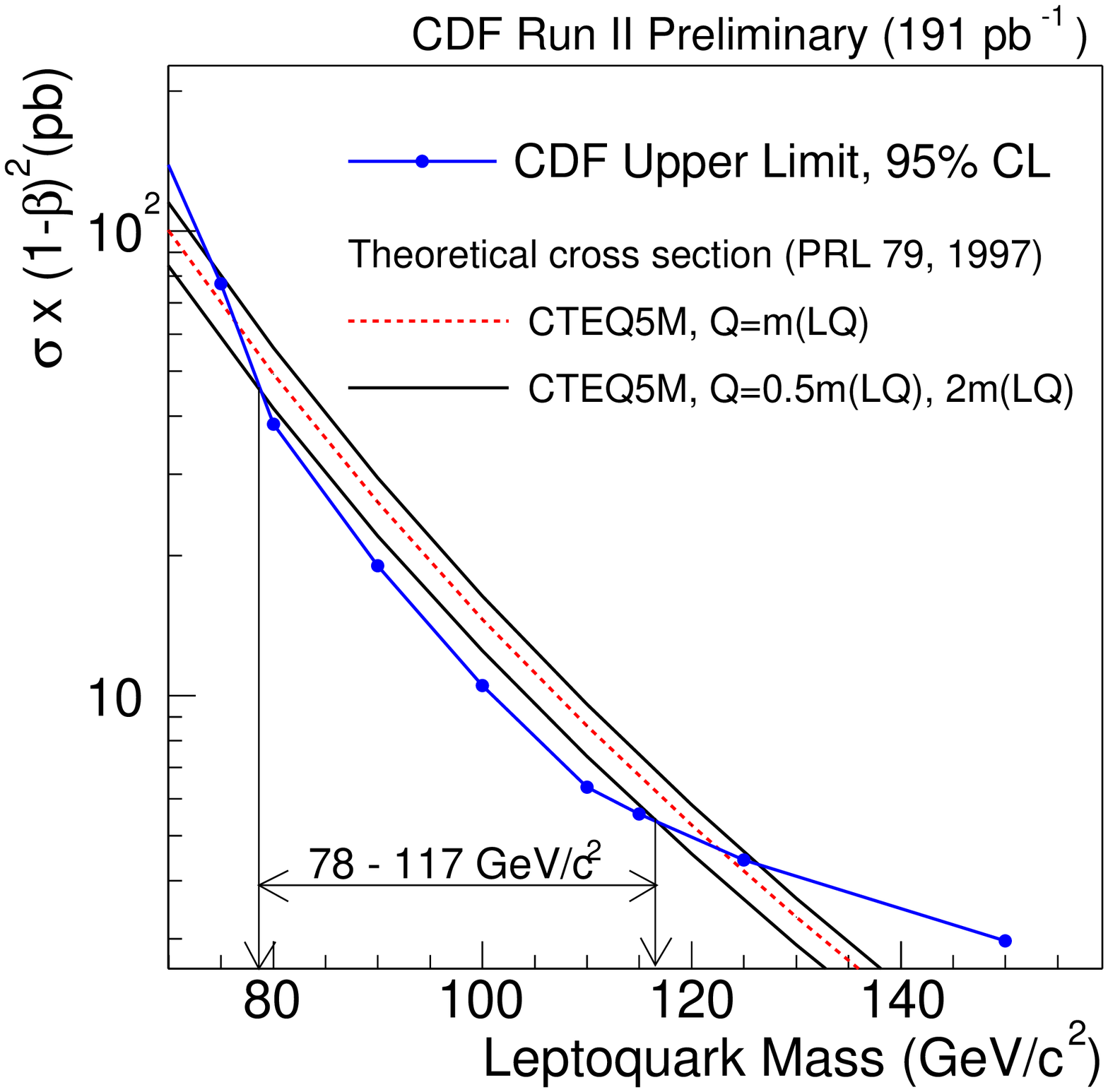,height=2.08in}
\caption{ CDF 95\% CL upper limit cross section times the squared
branching ratio for scalar $LQ$ pair production ($\beta$) = 0. }
\label{fig:LQA}
\end{center}
\end{minipage}
\end{figure}

 CDF has searched $LQ$ independent of generation in the $LQ\bar{LQ} \rightarrow
q\bar{q}\nu\bar{\nu}$ channel with 191 pb$^{-1}$ of data. The events selected
contain two or three jets and large \met ~\cite{ming}, with
different directions to reduce QCD background. Events with charged
leptons ($e$ or $\mu$) are vetoed to reduce contributions from W/Z+jets and $t\bar{t}$.
The number events expected from SM is 118 $\pm$ 14 events with 124 events are observed, consistent with the prediction. At 95\% C.L. a mass region from 78 \gvcc to 117 \gvcc 
is excluded, Figure~\ref{fig:LQA}.

\section{Summary}
The CDF and \dz collaborations are actively searching for new physics
beyond the SM using 200 pb$^{-1}$ collected from March 2002 to October
2003 (RunII). Although no evidence for new physics has been found so
far, the results have already improved those of Run I. The discovery
potential rises as the integrated luminosity increases, providing the
best opportunity for finding any evidence before LHC starts.


\section*{References}
\begin{thebibliography}{99}
\bibitem{CDF}  {\sc The CDF-II detector: Technical Design Report.}
By CDF-II Collaboration (R. Blair et al.). FERMILAB-PUB-96-390-E, Nov
1996. 234pp. and \\
CDF collaboration, F. Abe {\it et al.} \Journal{\NIM}{A}{271}{387}{1988}
\bibitem{D0} 

P.~M.~Tuts  [D0 collaboration],{\em Nucl. Phys. Proc. Suppl.}  {\bf 32}, 29 (1993)
and\\
D0 collaboration, S. Abachi {\it et al.} \Journal{\NIM}{A}{338}{185}{1994}
\bibitem{zp} M. Cvetic and S. Godfrey, hep-ph/9504216, 1995
\bibitem{zp1} F. Del Aguila, M. Quiros and F. Zwirner, \Journal{\NPB}{287}
{419} {1987}.
\bibitem{zp2} D. London and J.L. Rosner, \Journal{\PRD}{34}{5}{1530} {1986}
{095004}{2003}.

\bibitem{lh} N. Arkani-Hamed, A.G.Cohen and H.Georgi, \Journal{\PLB}{513}
{232}{2001}
\bibitem{lh2} T. Han, H.E. Logan, B. McElrath and  L.Wang,  \Journal{\PRD}{67}
\bibitem{lh3} N. Arkani-Hamed, A.G.Cohen , E. Katz and
A. E. Nelson,{\it hep-ph/0206021}. 

\bibitem{LED} JoAnne Hewett and Maria Spiropulu, hep-ph/0205106 (1998)
263
\bibitem{gled} G.F. Giudice, R. Ratazzi,J.D. Wells, \Journal{\NPB }{544}{3-38}
(1999)
 \bibitem{add}N. Arkani-Hamed, S. Dimopoulus, G. Dvali, \Journal{\PLB}{429}{263}{1998}
\bibitem{dxd}Salvatore Mele, Eur.Phys. J.C (2003).European Physical
Society, International Europhysics Conference on High Energy
Physics. EPS (July 17th-23rd 2003) in Aachen, Germany

\bibitem{jmet} \dz Collaboration, v.M. Abazov {\it et al.}. \Journal{\PRL}{90}{251802}{2003}.
\bibitem{RS} L.Randall, R.Sundrum, \Journal{\PRL} {83} {1999} {3370}.
\bibitem{exe} U.Baur, M.Spira and P.M Zerwas, \Journal{\PRD}{42}{3} {1990}

\bibitem{ja}C Jarlskog in {\em CP Violation}, ed. C Jarlskog
(World Scientific, Singapore, 1988).
\bibitem{tev}K.Dienes, E. Dudas, T. Ghergheta, \Journal{\NPB}{537}{47}{1999}.
	A. Pomarol, M. Quiros, \Journal{\PLB}{438}{255}{1999}
\bibitem{LQ}J.C. Pati, A.Salam, \Journal{\PRD}{19}{275}{1974} \\
	    E. Eitchen {\it et al.}, \Journal{\PRL}{50}{811}{1983}\\
	    J.L.Hewett, T.G.Rizzo, \Journal{\PRD}{183}{193}{1989}	


\bibitem{ming}
S.~M.~Wang  [CDF Collaboration],
``Search for Higgs, leptoquarks, and exotics at Tevatron,''
arXiv:hep-ex/0405075.
\end{thebibliography}
\end{document}